\documentclass{emulateapj}

\usepackage{natbib}

\shorttitle{X-ray emission from a diamond shock}
\shortauthors{Bonito et al.}

\begin{document}

\title{X-ray emission from protostellar jet HH 154: 
the first evidence of a diamond shock?}

\author{R. Bonito, and S. Orlando}
\affil{INAF -- Osservatorio Astronomico di Palermo, P.zza del Parlamento 1,
90134 Palermo, Italy}
\email{sbonito@astropa.unipa.it}

\author{M. Miceli\altaffilmark{1}, and G. Peres\altaffilmark{1}}
\affil{Dipartimento di Fisica-Specola Astronomica, Universit\`a di Palermo, P.zza del Parlamento 1, 90134 Palermo, Italy}

\author{G. Micela}
\affil{INAF -- Osservatorio Astronomico di Palermo, P.zza del Parlamento 1,
90134 Palermo, Italy}

\and

\author{F. Favata}
\affil{European Space Agency Community Coordination and Planning Office 8-10 rue Mario Nikis F-75738 Paris cedex 15 France}

\altaffiltext{1}{INAF -- Osservatorio Astronomico di Palermo, P.zza del Parlamento 1, 90134 Palermo, Italy}

\begin{abstract}
X-ray emission from about ten protostellar jets has been discovered and it appears as a feature common to the most energetic jets. Although X-ray emission seems to originate from shocks internal to jets, the mechanism forming these
shocks remains controversial. One of the best studied X-ray jet is HH 154 that has been observed by Chandra over a time base of about 10 years.  We analyze the
Chandra observations of HH 154 by investigating the evolution of its X-ray source. We show that the X-ray emission consists of a bright stationary component and a faint elongated component.  We interpret the observations by developing a hydrodynamic model describing a protostellar jet originating from a nozzle and compare the X-ray emission synthesized from the model with the X-ray observations. The model takes into account the thermal conduction
and radiative losses and shows that the jet/nozzle leads to the formation of a diamond shock at the nozzle exit. The shock is stationary over the period covered by our simulations and generates an X-ray source with luminosity and spectral characteristics in excellent agreement with the observations.
We conclude that the X-ray emission from HH 154 is consistent with a diamond shock originating from a nozzle through which the jet is launched into the ambient medium. We suggest that the physical origin of the nozzle could be related to the dense gas in which the HH 154 driving source is embedded
and/or to the magnetic field at the jet launching/collimation region.
\end{abstract}


\section{Introduction}
\label{Introduction}

Thanks to the capabilities of current X-ray observatories (Chandra and
XMM-Newton), a new class of X-ray sources has been discovered in the past
few years, namely those associated to protostellar jets. This class consists of
about ten members, HH 154 being one of the best studied. The X-ray source
associated with HH 154 has been previously observed by XMM-Newton
(\citealt{ffm02}) and by Chandra (\citealt{bfr03, fbm06}). These
observations had shown that the X-ray source is stationary over a period
of 4 years with a possible transient and faint component in the 2005
Chandra observations (\citealt{fbm06}).

Hydrodynamic models of the interaction between a continuously ejected
supersonic jet and the ambient medium predict X-ray emission caused by
shocks at the interaction front (\citealt{bop04,bop07}). X-ray emission
close to the base of the jet is predicted in the case of a pulsed-jet
(\citealt{bop10,bom10}). The pulsed-jet model is very effective in
reproducing the features and variability observed in most of the X-ray
emitting jets from low-mass young stellar objects (\citealt{bom10}). In
the pulsed-jet scenario, \citet{bom10} provided predictions on future
observations of HH 154 and claimed that a stationary X-ray source on a
period of $\approx 10$ years is unlikely. For a stationary X-ray source,
they suggested that a nozzle formed by either a dense medium or a magnetic
field may be at work at the launching site, leading to the formation of
a stationary diamond shock at the base of the jet.

The new Chandra observation of HH 154 (collected in 2009) provides a time base of 8 years to analyze the spectral and morphological variability of the X-ray source. 
Here we analyze the 2009 observations and compare the results of the data analysis with the previous Chandra (\citealt{bfr03, fbm06}) and XMM/Newton (\citealt{ffm02}) observations. We have developed a model describing a protostellar jet outflowing from a nozzle and compare the X-ray emission
synthesized from the model with the Chandra observations. Finally we interpret the new and the previous observations in the light of our model results. In Sect. 2 we describe the Chandra observations and their analysis; in Sect. 3 we discuss the hydrodynamic model of jet/nozzle and the comparison between model predictions and observations; in Sect. 4 we discuss the results and draw our conclusions.

\section{Chandra observations of HH 154}
\label{Chandra observations of HH 154}

We present the new Chandra/ACIS-S data of HH 154 (PI Schneider; ObsID 11016; $t_{exp} = 65.2$ ks) centered at (04:31:34.998, $+$18:07:51.95) (FK5), performed on 29 Dec. 2009, that provides $8$ years of time base between the first (2001, \citealt{ffm02}) and the last Chandra observations of HH 154, with the intermediate observation of 2005 (\citealt{fbm06}). To derive a uniform comparison between the three Chandra data-sets, we reprocessed all the data with the same method, using the latest CIAO 4.3 package, and filtered the data to restrict the energy to the $0.3 - 4.0$
keV band, as in \citet{fbm06}. Events were extracted for all observations from regions around the source and the background, near the position of the HH 154 jet (4:31:34.10, $+$18:08:04.9), following \citet{bfr03}. We defined a box $3.5''\times5''$ in the 2001 data-set, a box $3.5''\times5.5''$ in the 2005 data-set, and a box $3''\times6''$ in the 2009 data-set\footnote{The size of the boxes is different for different years to account for the variation of the morphology of the X-ray source.}, to extract the source events,
and four $10$ pixel radius circles for the background, from source-free parts of the image around the source. We extracted the spectra of the three data-sets and produced the ancillary response files, arf and rmf, by using SPECEXTRACT. We grouped the spectra to have a minimum of $5$ counts per bin. 

\begin{figure*}[!t]
\centerline{\includegraphics[angle=0,width=12cm]{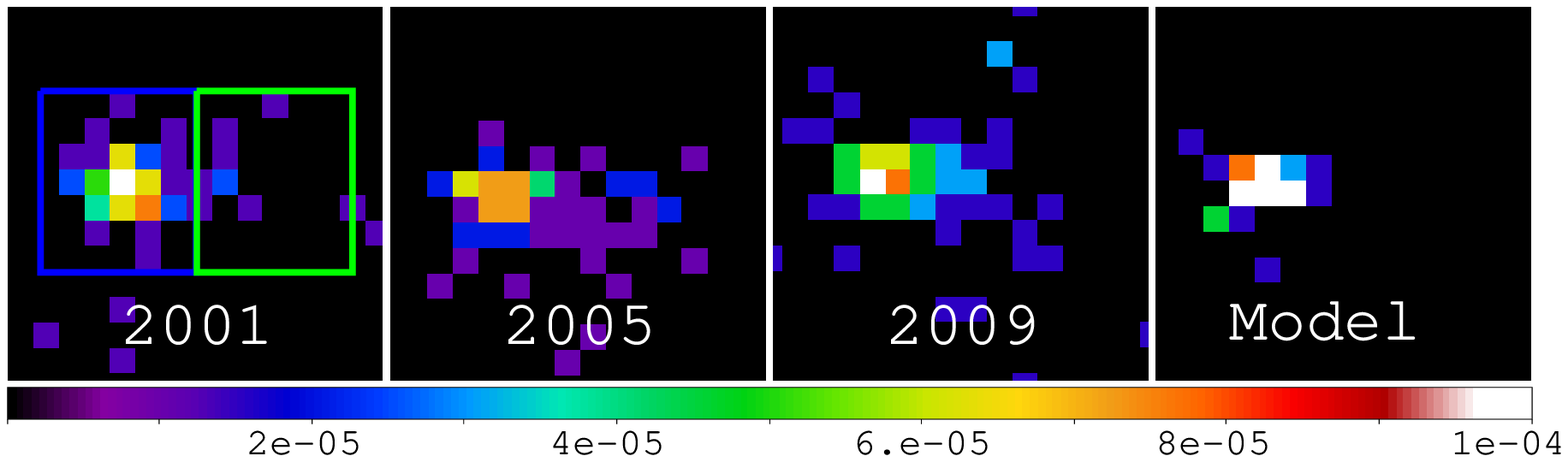}}
\centerline{\includegraphics[angle=0,width=12cm]{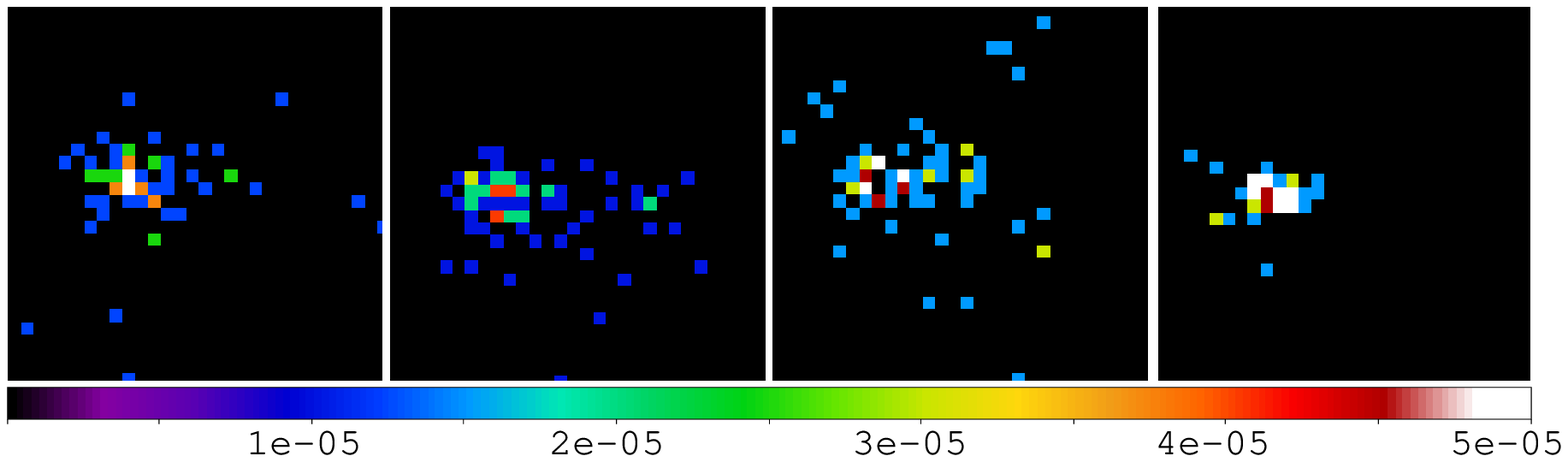}}
\centerline{\includegraphics[angle=0,width=12cm]{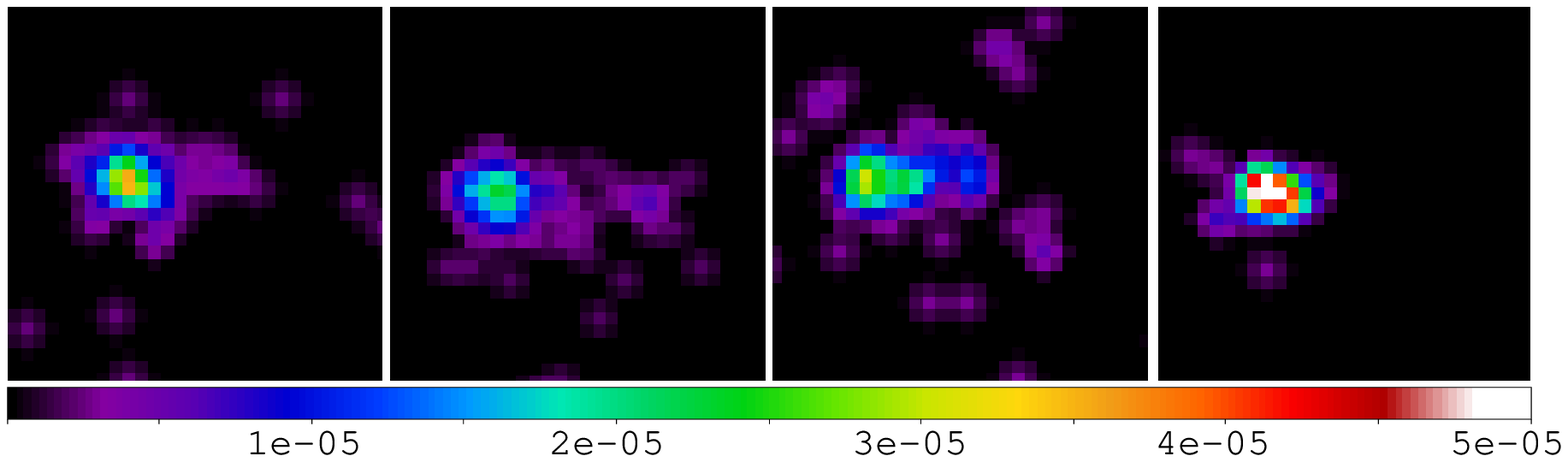}}
\caption{Upper panels: X-ray count rate maps ($0.3-4$ keV) of the 2001, 2005, and 2009 data-sets (first to third panels) at the native ACIS pixel size ($0.5''$). The last panel shows the synthetic X-ray map of the base of the jet as derived from the model at the same spatial resolution as Chandra/ACIS.
Central panels: The same as upper panels, but resampled at $0.25''$. EDSER technique has been applied on these data as explained in the text. Lower panels: The same as central panels, but with a smoothing applied on the images (Gaussian kernel of width $0.5''$). In each panel North is up and East is left. The angular size of each panel is $\approx7''\times7''$. The P.A. of the PSF asymmetry (see explanation in Sect. \ref{Results-obs}) is $93^{\circ}$ in 2001, $102^{\circ}$ in 2005, and $269^{\circ}$ in 2009. Note that this asymmetry can produce artificial extension on angular scales up to $1''$, as discussed in the text.}
\label{mappa-X-bin}
\end{figure*}

We also applied the sub-pixel repositioning algorithm available in CIAO 4.3 (EDSER) to the Chandra images to refine the event positions (\citealt{lkp04}). We resample the improved images at $0.25''$ pixel size, obtaining images with one-half of the native ACIS pixel scale. 
Fig. \ref{mappa-X-bin} shows the X-ray source associated with HH 154 in 2001 (first panels), 2005 (second panels), and 2009 (third panels), and the X-ray maps synthesized from our hydrodynamic model (discussed in Sect. \ref{The model}) with the same spatial resolution as ACIS (last panels). The upper panels show the images with the native ACIS pixel size ($0.5''$); the central panels show the images with sub-pixel repositioning algorithm applied and resampled with a pixel size $0.25''$; the lower panels show the improved resampled images smoothed using a Gaussian kernel of width $0.5''$.

\subsection{Results}
\label{Results-obs}

The three Chandra data-sets show that the X-ray emission is mainly located at the base of the HH 154 jet in all epochs, near the driving source. 
\citet{bfr03} found that the X-ray source is displaced by $0.5''-1''$ with respect to the L1551 IRS5 driving source.
The X-ray source consists of a bright knot which appears to be stationary in the time period covered by the observations (in the following the "stationary" component) and an elongated structure directed away from the driving source (in the following the "elongated" component) showing variability in the three data-sets (see Fig \ref{mappa-X-bin}). 
In particular, the latter component appears as a knot in images with EDSER applied (central and lower panels in Fig. \ref{mappa-X-bin}) which is much fainter than the knot of the stationary component. The position of the faint knot appears to be different in the three data-sets.
In particular, for the 2001 and 2005 data, we confirm the results of \citet{fbm06}, i.e. an elongation of the X-ray source corresponding to a proper motion of $0.7''$/yr approximately westward away from the driving source of the HH 154 jet. As for the 2009 observations, there is an hint of a faint knot closer to the brighter stationary source than in 2005. Note that recently an asymmetry in the Chandra PSF has been discovered\footnote{See \url{http://cxc.harvard.edu/ciao/caveats/psf\_artifact.html} for details.}, located at $P.A. = 195 -rollangle (\pm25)^{\circ}$, corresponding to $P.A. = 269.4^{\circ}$ in 2009 observations, i.e. roughly the direction of the extension of the X-ray source detected in HH 154. 
We checked if this instrumental effect may influence the observed morphology of the X-ray source and found that the asymmetry of the PSF does not affect our images on scales larger than 1 arcsec. The elongated structure visible in the images, therefore, is not an artifact of the instrument.
The maximum length of the whole X-ray source is $\approx5''$ ($\approx700$ AU at $D=140$ pc).

\begin{figure}[!t]
\centerline{\includegraphics[angle=-90,width=9cm]{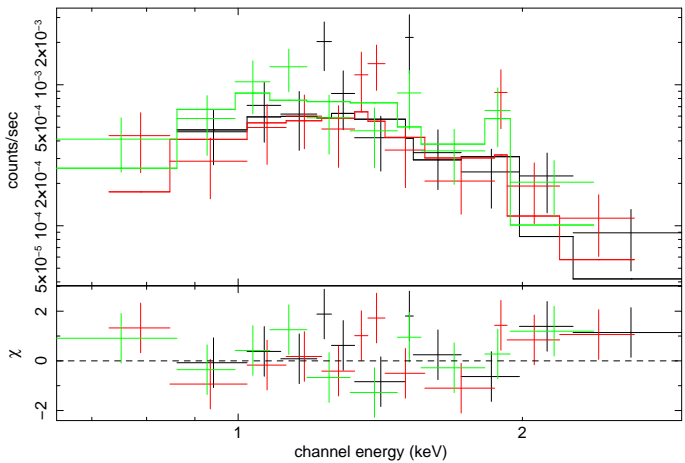}}
\caption{Best-fit X-ray spectral model superimposed on the three Chandra data-sets (2001 in black, 2005 in red, and 2009 in green) fitted simultaneously.}
\label{spettri}
\end{figure}

We performed a spectral analysis of the individual data sets\footnote{We have verified that our procedure reproduces the results obtained by \citet{bfr03} and \citet{fbm06}.}. All spectra  are well fitted by an
absorbed thermal plasma (APEC in XSPEC). The count rates of the three observations ($0.76\pm0.10$ cnts/ks in 2001, $0.65\pm0.08$ cnts/ks in 2005, and $0.89\pm0.12$ cnts/ks in 2009) are compatible within the Poisson error, indicating that the X-ray luminosity is constant. Moreover the source shows no spectral variability in terms of temperature ($T$) and emission measure ($EM$) in the timescale analyzed here. We have thus fit the three data-sets simultaneously finding the values reported in Table \ref{fit}.

\begin{table*}
\begin{center}
\caption{Best-fit values derived from the three data-sets simultaneously fitted.\label{fit}}
\begin{tabular}{crrrrrrrrrrr}
\tableline\tableline
$T$ & $EM$ & $F_{X}$ & $L_{X}$ & $N_{\rm H}$ \\
(K) & (cm$^{-3}$) & (erg cm$^{-2}$ s$^{-1}$) & (erg s$^{-1}$) & (cm$^{-2}$) \\
\tableline
$7_{-2}^{+8}\times10^{6}$ &$5_{-3}^{+4}\times10^{51}$ &$5_{-3}^{+4}\times10^{-14}$ &$1.3_{-0.9}^{+0.8}\times10^{29}$  &$1.2\pm0.2\times 10^{22}$ \\
\tableline
\end{tabular}
\end{center}
\end{table*}

The $N_{\rm H}$ is well constrained (more than in the analysis of XMM-Newton data; \citealt{ffm02}) by fitting the three data-sets simultaneously but it cannot be constrained by the individual data-sets. In the latter case we fixed $N_{\rm H}$ to the value derived in the simultaneous fitting of the three data-sets. 
The three data-sets and the best-fit model are shown in Fig. \ref{spettri}.

To investigate possible spatial variations of the spectral properties, we selected two regions, confining the stationary component (blue box in Fig. \ref{mappa-X-bin}) and the elongated component (green box in Fig. \ref{mappa-X-bin}), and computed the median photon energy, $MPE$, within each region in 2001, 2005, and 2009. $MPE$ has been proved to be a robust indicator of the spectral properties of a source in the case of low statistics (\citealt{hsg04}). In each data-set the total number of counts considered is $\approx50$, in the stationary component, and $\sim10$, in the elongated component. The stationary component shows no temporal variability in the three epochs, its MPE varying in the range $MPE_{\rm s} = 1.35
- 1.42$ keV. The median energy of the elongated component $MPE_{\rm e}$ is always lower than $MPE_{\rm s}$. By considering the three observations altogether $MPE_{\rm s} \approx 1.4$ keV, while the elongated component appears softer with $MPE_{\rm e} \approx 1.0$ keV. 

This result is supported by the Kolmogorov-Smirnov test we performed to check if the photon energy distributions of the two regions differ significantly and vary with time. We found that the stationary and the elongated component are compatible with being constant over the time base analyzed. We also found that the two components are dissimilar at the $99.99\%$ confidence level. In the light of this result, we have fitted simultaneously the data collected in the three epochs for each of the two components and found that the stationary component is described by a plasma with $T = (7 -14)\times10^{6}$ K and the elongated component by a plasma with $T < 7\times10^{6}$ K (assuming the same $N_{H}$ for the two spectra), confirming that the elongated component is softer.

\section{The model}
\label{The model}

X-ray observations of HH 154 suggest the presence of a steady shock at
the base of the jet over a time interval of about $8$ years. Examples
of quasi-stationary shocks in jets are the diamond shocks, namely
stationary shock patterns appearing in supersonic jets when the jet
material outflowing from a nozzle is slightly over or under-expanded or,
in other words, when the pressure of the gas exiting the nozzle is below
or above the pressure of the ambient medium. \citet{bom10} suggested that
a diamond shock forming near the launching/collimation site of the jet is
the most likely mechanism leading to a stationary X-ray source in HH 154.

\begin{figure}[!t]
\centerline{\includegraphics[angle=0,width=8cm]{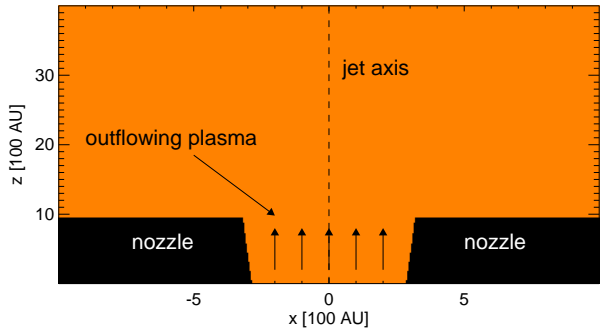}}
\caption{Enlargement of the nozzle through which the jet is ejected into
the ambient medium.}
\label{nozzle}
\end{figure}

To test the above idea, we extended the hydrodynamic model of
\citet{bop07} (to which the reader is referred for more details),
describing the interaction between a continuously ejected supersonic jet
with the unperturbed medium, by adding a diverging nozzle at the base
of the jet included as an impenetrable body (see Fig. \ref{nozzle}). The
model is described by the hydrodynamic equations solved in two dimensions
(adopting cylindrical coordinates and assuming axisymmetry with the jet
axis coincident with the axis of symmetry; \citealt{bop07}) and takes
into account the thermal conduction (including heat flux saturation)
and radiative losses from optically thin plasma. The calculations were
performed using FLASH, a well tested adaptive mesh refinement multiphysics
code (\citealt{for00}).

The jet with temperature $T_{\rm j} = 10^{4}$ K outflows from the nozzle
and propagates through an initial uniform ambient medium with the
same temperature. The initial jet radius depends on the nozzle size,
chosen accordingly with observations of the order of tens of AU (see
\citealt{bfr03}). We have explored a wide space of model parameters
defined by the particle number density of the jet ranging between
$n_{\rm j}= 3\times 10^2$ and $5\times 10^3$ cm$^{-3}$, the density of
the ambient medium ranging between $n_{\rm a}= 3\times 10^3$ and $10^5$
cm$^{-3}$, the throat of the nozzle with radius ranging between
$R_{\rm th} = 15$ and 200 AU, and the jet velocity ranging between $u_{\rm j} = 1000$ and $1500$ km s$^{-1}$. We assume that the direction of propagation of the jet is perpendicular to the line of sight.
Our best-fit model is characterized by $n_{\rm j}= 3\times 10^2$ cm$^{-3}$, $n_{\rm a}= 3\times 10^3$ cm$^{-3}$, $u_{\rm j} = 1500$ km s$^{-1}$, and $R_{\rm th} = 100$ AU. The jet is overexpanded and travels through an initially denser ambient medium (light jet scenario; see \citealt{bop07}) with ambient-to-jet density contrast $\nu = n_{a}/n_{j} = 10$.

The computational domain is $(1000\times4000)$ AU ($\approx7\times27$
arcsec at $D = 140$ pc). Inflow boundary conditions are imposed
at $z = 0$ and $r < R_{\rm th}$, axisymmetric boundary conditions are
imposed along the jet axis (consistent with the adopted symmetry), and
outflow boundary conditions are assumed elsewhere.  The maximum spatial
resolution achieved by our simulations is $\approx 2$ AU, using
$5$ refinement levels with the PARAMESH library (\citealt{mom00}), which
corresponds to covering the initial jet radius of our best-fit model
with 50 grid points. The grid resolution is increased inside the nozzle
and around the nozzle exit to capture the diamond structure forming
there.
Note that the pixel size of the hydrodynamic model at its maximum spatial resolution ($2$ AU) corresponds to a spatial resolution of $0.014''$ at the distance of $140$ pc, i.e. much lower than the nominal spatial resolution of ACIS ($0.5''$).
As discussed below, this implies that the synthetic maps derived from the hydrodynamic model needs to be rebinned to match the ACIS pixel size in order to compare the model results with the observations.

\subsection{Synthesis of X-ray emission}
\label{Synthesis of X-ray emission}

Following \cite{bop07}, we synthesized the X-ray emission associated with the jet from the model results, recovering the 3-dimensional spatial
distributions of mass density and temperature.
We then derived the emission measure, $EM$, and temperature, $T$, for each fluid element and the distribution $EM(T)$ integrated along the line of sight, for each element, in the temperature range $10^{3}-10^{8}$ K (using $75$ bins equispaced in $\log T$). From the EM(T), we synthesize the maps of the X-ray emission and the focal plane spectra using the APEC spectral code, and considering photon count statistics comparable with that of the
observations. The source is assumed to be at a distance $D=140$ pc. We filtered the emission through the Chandra/ACIS instrumental response and the interstellar column density at the best-fit value $N_{H}=1.2\times 10^{22}$ cm$^{-2}$ (see Sect. \ref{Results-obs}).
In particular, to compare the synthesized images with the different observations, we have used the rmf and arf files generated for the three data-sets (2001, 2005, and 2009) to account for the correction to the charge transfer inefficiency and the degradation in low-energy response due to contaminant buildup on the optical blocking filter.
  
To directly compare the synthesized images with the observations, we rebinned the model images (whose spatial resolution is $2$ AU, corresponding to $0.014''$ at $140$ pc) so as to have the same bin size as the Chandra images as shown in Fig. \ref{mappa-X-bin} ($0.25''$, or $0.5''$).

We also convolved the synthesized X-ray image with the specific PSF, created at the proper energy for each data set, by using the Chandra standard analysis tools. We finally included Poisson fluctuations to mimic the photon count statistics.

\subsection{Results}
\label{Results-model}

\begin{figure*}[!t]
\centerline{\includegraphics[angle=0,width=11cm]{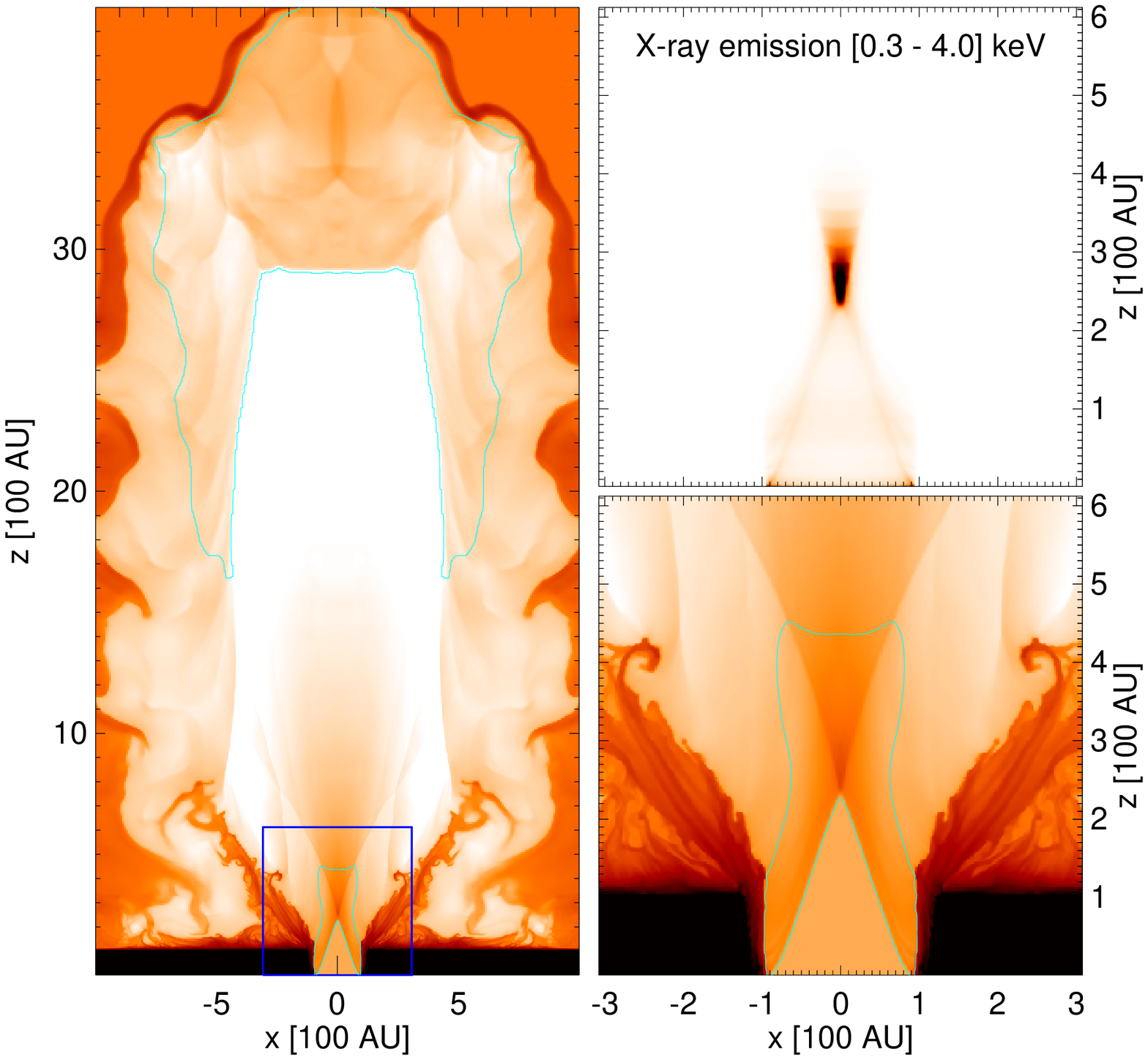}\hspace{1.8cm}\includegraphics[angle=0,width=0.7cm]{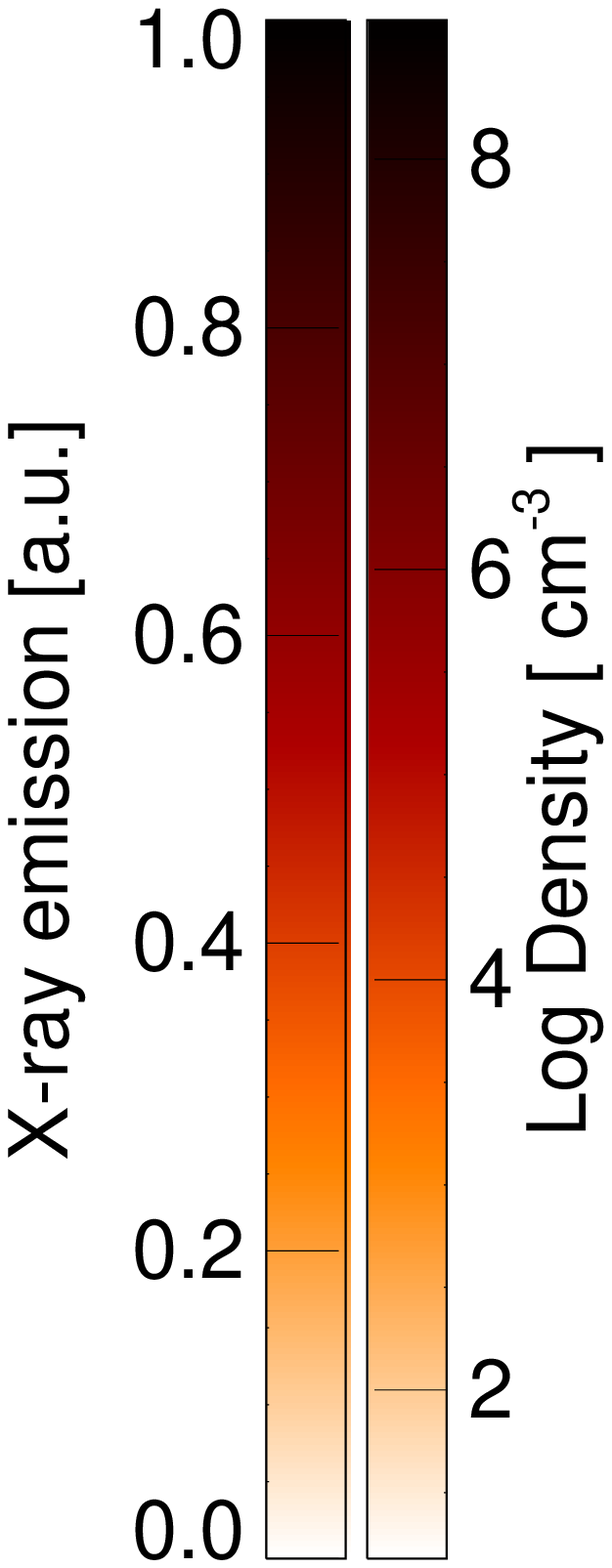}}
\vspace{1cm}
\caption{Density map (left panel) with the $2\times10^{6}$ K contour
superimposed, an enlargement of the base of the computational domain
(lower panel on the right), and the X-ray map synthesized from the model
(upper panel on the right) at the maximum spatial resolution achieved
from the simulation. $100$ AU correspond to $\approx 0.7''$ at $140$ pc.}
\label{mappa}
\end{figure*}

Figure~\ref{mappa} shows the spatial distribution of density in a
slice in the $(x,z)$ plane (left panel) for the best-fit model at time
$t\approx120$ yrs and a close up view of the nozzle site (right bottom
panel). After the jet exits the nozzle, it propagates through the ambient
medium forming a shock with temperature $T\approx 10^7$ K at its head
and a cold cocoon enveloping it. The cocoon is characterized by low
temperatures ($T < 10^5$ K), is dominated by the radiative cooling,
and is strongly perturbed by the hydrodynamic instability developing
there as the jet progresses through the medium, the thermal conduction
being inefficient in damping the instability (see \citealt{bom10,bop10}).

The nozzle determines a diamond-shaped shock past the nozzle exit with
peak temperature $T\approx 8\times 10^6$ K (see right bottom panel in
Fig.~\ref{mappa}). This diamond structure has its origin inside the
nozzle and appears as a shock emerging from the nozzle and reflecting
just past of the nozzle exit. After its formation ($t\approx 50$ yrs),
the diamond shock is almost stationary until the end of the simulation
for $\approx 100$ yrs. The thermal conduction is rather efficient in
the post-shock region given the high temperatures there ($T> 10^6$
K) and is crucial in stabilizing the diamond structure, damping the
hydrodynamic instability developing past the nozzle exit. Auxiliary
simulations performed without the thermal conduction have shown that the
diamond shock would be unstable if the thermal conduction is neglected,
the hydrodynamic instability heavily perturbing the flow structure at
the nozzle exit.

Analyzing the X-ray emission synthesized from the hydrodynamic model, as described in Sect. \ref{Synthesis of X-ray emission}, we investigated both the morphology and spectral properties of the synthetic X-ray sources.
The X-ray emission from the modeled jet consists of two main features:
a quasi-stationary source associated with the diamond shock at the
jet base and a moving source associated with the shock at the head
of the jet. The latter is a transient feature we are not interested
in\footnote{The X-ray emission at the head of the jet forms because we
simulate the early evolution of the jet when its head is close to the
stellar driving source. We expect that the head of the jet progressively
cools down because of radiative cooling as it goes away from the driving
source, eventually emitting mainly in the optical band.} and does not
influence the evolution of the diamond shock at the base of the jet;
therefore we will not discuss its properties in the following.

The X-ray luminosity of the diamond shock is $L_{\rm X}\approx 5\times
10^{29}$ erg and is stationary over $\approx 100$ yrs. This value
is similar to that observed for HH 154 that is almost stationary in
about $8$ years. 

By comparing the total flux derived from the model with the specific rmf and arf response of each data-set, we have verified that the degraded QE of the instrument in the time baseline analyzed affects the synthesized count rate for less than $7\%$. This confirms that the source flux can be assumed constant over $8$ yrs, within the Poisson errors.

\begin{figure}[!t]
\centerline{\includegraphics[angle=0,width=8cm]{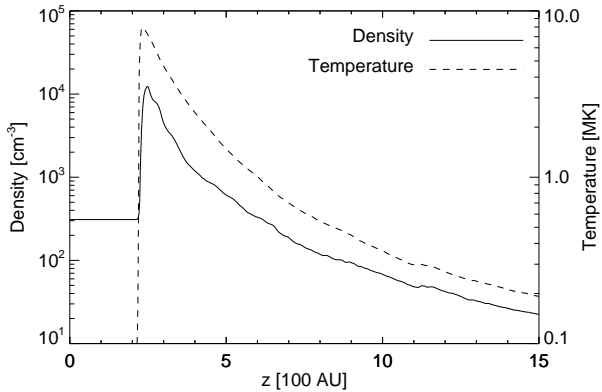}}
\caption{Profiles of particle number density (solid line) and temperature
(dashed line) along the jet axis in the region where the diamond shock
forms.}
\label{mod_prof}
\end{figure}

The right upper panel in Fig. \ref{mappa} shows the
synthetic X-ray emission arising from the shock integrated along the
line-of-sight. Most of the emission originates just behind the shock
in a bright and compact knot with temperature $T\approx 8\times 10^6$
K. The knot is surrounded by a diffuse region elongated along the jet
axis, characterized by lower temperatures ($T\approx 1-2\times 10^6$
K). Figure~\ref{mod_prof} shows the profiles of density and temperature
along the jet axis in the region where the diamond shock forms.

We found that the spectrum synthesized from the hydrodynamic model, as explained in Sect. \ref{Synthesis of X-ray emission}, can be fitted with
one isothermal component which is compatible with that derived from
the three data-sets of Chandra. 

We rescaled the synthetic X-ray image shown in Fig.~\ref{mappa} to the
Chandra/ACIS pixel size (last panels in Fig. \ref{mappa-X-bin}). The emission within the nozzle is assumed to be
totally absorbed. We found that the spatial scales of the X-ray emitting
diamond shock at the same spatial resolution of Chandra are consistent
with the size of the HH 154 X-ray emitting source: a synthetic X-ray
source of a few arcsec at the base of the jet consisting of a bright
point-like component surrounded by a faint and elongated component along
the jet axis.

\begin{figure}[!t]
\centerline{\includegraphics[angle=0,width=8cm]{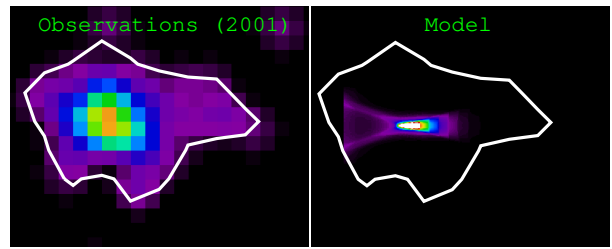}}
\caption{Smoothed 2001 image of the HH 154 X-ray source (left panel) with a pixel size of $0.25''$ compared with the high resolution ($0.014''$) image derived from the model (right panel). The contour of the observed source is superimposed on the images.
Since we verified that the direction of the PSF asymmetry in 2001 is along the counter jet and since this asymmetry can produce artificial structures only out to $1''$ (four pixels at the $0.25''$ pixel size of the image on the left panel), the observed elongation is not an artifact due to the PSF asymmetry. The angular size of each panel is $\approx 5''\times 4''$. In both panels, North is up and East is left.}
\label{contour}
\end{figure}

In Fig. \ref{contour} we compare the smoothed 2001 image with a bin size $0.25''$ (left panel) with the X-ray source derived from the model at its maximum spatial resolution, $0.014''$, (right panel). The 2001 image contour is superimposed on the modeled source.

\section{Discussion and conclusions}
\label{Discussion and conclusions}

The analysis of the observations of HH 154 in three
different epochs with Chandra reveals a faint and elongated X-ray
source displaced by $0.5-1$ arcsec (\citealt{bfr03}) from the L1551 IRS\,5 protostar (the
driving source of the jet) along the jet axis. The source appears to be
quasi-stationary over a time base of $\approx 8$ yrs without appreciable
proper motion and variability of X-ray luminosity and of temperature. The
morphological analysis shows that the X-ray source consists of a bright
stationary component with temperature $T > 7\times
10^{6}$ K surrounded by an elongated cooler component extended in
the direction away from the driving source, with temperatures $T < 7 \times 10^{6}$ K. 
Very recently \citet{sgs11} analyzed the same data-sets finding similar observational results in terms of X-ray luminosity, spectral parameters, and morphology, independently showing the robustness of the derived parameters that form the basis of our comparison with a simulation of the jet based on detailed hydrodynamic models.

As shown in \citet{bff08}, the X-ray source is not perfectly aligned with the optical jet observed in HH 154 (see Fig. 13 in \citealt{bff08}). In fact the HST images of \citet{fld05} show that the optical jet from HH 154 is along $P.A.\approx254^{\circ}$ (see also \citealt{phk02}), while from the X-ray data we derive a $P.A.\approx270^{\circ}$. \citet{bom10} suggested that an ejection direction varying in time could explain the misalignment between the X-ray source and the optical jet. 
Since the jet driving source, L1551 IRS5, is known to be a binary system (\citealt{bc85}), a jet precession could be induced due to the presence of the companion star. 

The absorption column density derived from the analysis of the three data-sets is too low if compared with the 150 mag of absorption of L1551 IRS\,5, confirming the results of \cite{bfr03} and \cite{fbm06}. This fact together with the evident displacement of $0''.5 - 1''$ of the source from L1551 IRS\,5 and the lack of temporal variation in the X-ray flux and spectral properties suggest that the X-ray emission detected in the three epochs unambiguously arises from the jet and cannot be of stellar origin.

The observations suggest therefore that the X-ray emission of HH
154 originates in a standing shock located at the base of the jet.
\citet{bom10}, by analyzing the X-ray emission arising from a pulsed
jet model, have discussed the possibility to produce a standing shock at
the base of the jet as a result of multiple self-interactions of plasma
blobs ejected in different epochs by the driving source and concluded
that this mechanism is unlikely. In addition these authors suggested
that the most likely mechanism leading to a standing shock over a period $>5$ yrs might be a diamond shock forming near the launching/collimation site of the jet.

This idea is expanded here by developing a model of jet outflowing
through a nozzle. We found that, in such a configuration, a standing
diamond shock forms just past the nozzle exit at the base of the jet. The
shock is stabilized under the action of the thermal conduction which
damps the hydrodynamic instability developing within the cocoon and
heavily perturbing the flow structure. We found that the X-ray emission
arising from the diamond shock has morphology and spectral characteristics
in excellent agreement with those derived from the three data-sets of
HH 154. The model also predicts that the X-ray emitting plasma of the
diamond shock cools down at larger distances from the driving source as
inferred from the observations. In fact, from the Chandra data,
we found that the MPE increases toward the base of the jet. While we
cannot rule out that this result is associated with variations in $N_H$
(\citealt{fld05} found an increasing absorption towards the driving
source along the jet axis), higher values of MPE can be indicative of
higher temperatures. Such variations of plasma temperature would be naturally explained by our model.
We conclude therefore that HH 154 offers the first evidence of a
standing diamond shock at the base of the jet probably near the jet
launching/collimation region.

We can infer a characteristic size, $l_{sh}$, of the X-ray emitting source from the spectral analysis, using the value of the $EM_{best-fit}$, and from the hydrodynamic model results, using the maximum particle density value derived from the diamond shock modeled. In particular, we find $l_{sh} > V^{1/3}=4\times10^{14}$ cm, where $V$ is the volume derived from the $EM = 0.8 n^{2} V$ (following \citealt{ffm02}) and $n_{MAX}\approx10^{4}$ cm$^{-3}$ (see the peak of density in Fig. \ref{mod_prof}) at $140$ pc. This value is in good agreement with both the observed radius of HH 154, $r_{j}\approx30$ AU (see discussion on this parameter in both models and observations in \citealt{bop07} and \citealt{bff08}), and with the size of the diamond shock modeled.
Therefore, although the spatial resolution of the ACIS data, improved by sub-pixel techniques, is more than an order of magnitude lower than that achieved by our numerical model, its diagnostic power allows us to infer detailed information on physical scales comparable to numerical simulations.

The data, after this analysis, clearly show the presence of an elongated structure on the right side of the main source. Concerning an eventual evolution of this elongated structure the most conservative interpretation could be that it is steady and we can hardly constrain its features, due to the limited photon statistics. However, given the insight provided by the model and the evidence that stellar jets flows are inherently variable, another interpretation is that we are observing the diamond shock variability and possibly knots formation and motion due to the changes of the jet flow (cf. \citealt{bom10} for example of effects of variable jet flows).
The sequence of the smoothed images of HH 154 with a spatial scale of $0.25''$ shown in Fig. \ref{mappa-X-bin} suggests the presence of subsequent knots with a detectable proper motion. 
In particular, by comparing the 2001 and 2005 data-sets, we confirm the results of \citet{fbm06} who found a detectable westward proper motion of the elongated component of the X-ray source, corresponding to $\approx 500$ km/s; we find a hint of elongation again westward in the 2009 observations away from the jet driving source, but closer to the stationary source than in the 2005 observations. This evidence may suggest the presence of a newly formed knot propagating away from the diamond shock. However note that, in the 2009 observation, the PSF asymmetry is directed almost along the X-ray extension axis, and therefore it could influence the X-ray source elongation up to an angular scale $\approx1''$.
Therefore, both the evidence of a standing shock at the base of the HH 154 jet over a $8$ yrs timebase and a moving knot in 2005 together with a hint of a new elongation in the 2009 smoothed image, indicate the scenario of a nozzle, creating the standing shock, in the presence of a pulsed jet, as described in \citet{bom10}, which may account for the moving/elongated component.
No matter how one interprets the observations, there is a clear need for future observations of HH 154.

The physical origin of the nozzle could be related to the dense
gas in which the L1551 IRS\,5 protostar is embedded and/or the intense
stellar magnetic field at the jet launching/collimation region. In the
hypothesis that a magnetic nozzle causes the diamond shock observed
in HH 154, the Chandra observations and the comparison with our model
offer the possibility to constrain the magnetic field strength near
the jet launching/collimation region. In fact, the model provides the
total plasma pressure $(p_{\rm sh}+\rho_{\rm sh}u_{\rm sh}^2)$ in the
X-ray emitting diamond shock that reproduces the observations, where
$p_{\rm sh}$, $\rho_{\rm sh}$ and $u_{\rm sh}$ are the pressure, mass
density and velocity in the post-shock region close to the nozzle exit,
respectively. Then, assuming the plasma $\beta = (p_{\rm sh}+\rho_{\rm
sh}u_{\rm sh}^2)/(B^2/8\pi) \approx 1$, where $B$ is the magnetic field
strength, we derive $B\approx 5$ mG in the magnetic nozzle at the base
of the jet.  Interestingly, this value is consistent with that inferred
by \cite{bfr03}, namely $B = 1-4$ mG, in the context of shocks
associated to jet collimation, and by \citet{sgs11}, who find $B\approx6$ mG, which is a reasonable value at the jet basis near the driving source, according to \citet{hfv07}. 
We suggest therefore that the comparison
between our model and the X-ray observations of HH 154 may allow us to
probe the launching/collimation region near the driving source, very
difficult to be directly observed in systems so obscured.

\begin{acknowledgements}

We thank an anonymous referee for constructive and helpful comments.
R. B. would like to thank Dr. M. Caramazza, Dr. E. Flaccomio, and Dr. G. Sacco  for interesting discussions about the CIAO tools. 
This research has made use of data obtained from the Chandra Data Archive, and software provided by the Chandra X-ray Center (CXC) in the application package CIAO.
The software used in this work was in part developed by the DOE-supported
ASC/Alliances Center for Astrophysical Thermonuclear Flashes at
the University of Chicago. The calculations were performed at the SCAN
(Sistema di Calcolo per l'Astrofisica Numerica) HPC facility of the INAF
-- Osservatorio Astronomico di Palermo. This work was supported in part
by Agenzia Spaziale Italiana under contract No. ASI-INAF (I/009/10/0).

\end{acknowledgements}


\end{document}